\documentclass[camera,hyphens]{jpaper}
\usepackage{cite}
\usepackage{amsmath,amssymb,amsfonts}
\usepackage{graphicx}
\usepackage{fancyhdr}
\usepackage{hyperref}
\usepackage{url}
\usepackage{color}
\usepackage[dvipsnames]{xcolor}
\usepackage{tikz}
\usepackage[ddmmyyyy]{datetime} 
\usepackage{algorithm}
\usepackage{comment}
\usepackage[noend]{algpseudocode}
\usepackage{setspace}
\usepackage{xcolor}
\usepackage{balance}
\makeatletter
\algrenewcommand\ALG@beginalgorithmic{\footnotesize}
\makeatother

\algrenewcommand{\algorithmiccomment}[1]{\hfill$\triangleright$ \textit{#1}}
\algrenewcommand\algorithmicindent{1.0em}

\algnewcommand{\IIf}[1]{\State\algorithmicif\ #1\ \algorithmicthen}
\algnewcommand{\EndIIf}{\unskip\ \ }

\usetikzlibrary{positioning}

\definecolor{blue(pigment)}{rgb}{0.2, 0.2, 0.7}
\definecolor{dgreen}{rgb}{0.00, 0.75, 0.00}
\definecolor{ddgreen}{rgb}{0.00, 0.50, 0.00}
\definecolor{ddred}{rgb}{0.50, 0.00, 0.00}

\def\BibTeX{{\rm B\kern-.05em{\sc i\kern-.025em b}\kern-.08em
    T\kern-.1667em\lower.7ex\hbox{E}\kern-.125emX}}

\definecolor{MidnightBlue}{rgb}{0.1, 0.1, 0.44}
\newcommand{\versionnum}[0]{3.5~---~\today~@~\currenttime~CET} 
\newif\ifcameraready
\camerareadytrue
\fancypagestyle{firstpage}{
  \fancyhf{}

  \fancyfoot[C]{\thepage}
}  

\definecolor{mGreen}{rgb}{0,0.6,0}
\definecolor{mGray}{rgb}{0.5,0.5,0.5}
\definecolor{mPurple}{rgb}{0.58,0,0.82}
\definecolor{backgroundColour}{rgb}{0.95,0.95,0.92}
\definecolor{gray97}{gray}{0.97}

\newcommand{\di}[1]{{\color{black}#1}}

\title{Exploiting Near-Data Processing \\to Accelerate Time Series Analysis} 
\newcommand{\affilUMA}[0]{\textsuperscript{\S}}
\newcommand{\affilETH}[0]{\textsuperscript{$\ddagger$}}
\newcommand{\affilNTUA}[0]{\textsuperscript{$\dagger$}}

\author{
{Ivan Fernandez\affilUMA}\qquad~~~%
{Ricardo Quislant\affilUMA}\qquad~~~%
{Christina Giannoula\affilNTUA}\qquad~~~%
\vspace{2pt}
{Mohammed Alser\affilETH}\\
{Juan Gómez-Luna\affilETH}\qquad~~~%
{Eladio Gutiérrez\affilUMA}\qquad~~~%
{Oscar Plata\affilUMA}\qquad~~~%
\vspace{6pt}
{Onur Mutlu\affilETH}\\% 
\emph{{\affilUMA University of Malaga \qquad  \qquad \affilNTUA National Technical University of Athens \qquad \affilETH ETH Z{\"u}rich
}}%
}
\begin{document}
\bstctlcite{IEEEexample:BSTcontrol}

\maketitle

\fancyhead{}
\ifcameraready
 \thispagestyle{plain}
 \pagestyle{plain}
\else
 \fancyhead[C]{\textcolor{MidnightBlue}{\emph{ICCD 2020 Camera Ready Version \versionnum}}}%\today, \ampmtime}}}
 \fancypagestyle{firststyle}
 {
   \fancyhead[C]{\textcolor{MidnightBlue}{\emph{ICCD 2020 Camera Ready Version \versionnum}}}%%\today, \ampmtime}}}
   \fancyfoot[C]{\thepage}
 }
 \thispagestyle{firststyle}
 \pagestyle{firststyle}
\fi

\begin{abstract}

A time series is a chronologically ordered set of samples of a real-valued variable that can contain millions of observations. Time series analysis is used to analyze information in a wide variety of domains~\cite{SS17}: epidemiology, genomics, neuroscience, medicine, environmental sciences, economics, and more. Time series analysis includes finding similarities (\emph{motifs}) and anomalies (\emph{discords}) between every two subsequences (i.e., slices of consecutive data points) of the time series. There are two major approaches for motif and discord discovery: approximate and exact algorithms. Approximate algorithms are faster than exact algorithms, but they can provide inaccurate results or limited discord detection, which cannot be tolerated by many applications (e.g., vehicle safety systems). Unlike approximate algorithms, exact algorithms do not yield false positives or discordant dismissals, but can be very time-consuming on large time series data. Thus, \emph{anytime} versions (aka interruptible algorithms) of exact algorithms are proposed to provide approximate solutions quickly and can return a valid result even if the user stops their execution early. The state-of-the-art exact \emph{anytime} method for motif and discord discovery is {\it matrix profile}~\cite{MPROFILEI}, which is based on Euclidean distances and floating-point arithmetic. We evaluate a recent CPU implementation of the \emph{\di{matrix profile}} algorithm~\cite{MPROFILEXI} on a real multi-core machine (Intel Xeon Phi KNL~\cite{INTELKNL}) and observe that its performance is heavily bottlenecked by data movement. In other words, the amount of computation per data access is not enough to hide the memory latency and thus time series analysis is  memory-bound. This overhead caused by data movement limits the potential benefits of acceleration efforts that do not alleviate the data movement bottleneck in current time series applications. 
\vspace{-1mm}

Several CPU and GPU implementations of \emph{\di{matrix profile}} have been proposed in the literature. However, these acceleration efforts still require transferring the time series data from the main memory to the CPU/GPU cores, leading to the data movement bottleneck. 
Near-Data Processing (NDP)~\cite{stone1970logic, Kautz1969, shaw1981non, kogge1994, gokhale1995processing, patterson1997case, oskin1998active, kang1999flexram, Mai:2000:SMM:339647.339673,murphy2001characterization, Draper:2002:ADP:514191.514197,aga.hpca17,eckert2018neural,fujiki2019duality,kang.icassp14,seshadri.micro17,seshadri.arxiv16,Seshadri:2015:ANDOR,seshadri2013rowclone,angizi2019graphide,kim.hpca18,kim.hpca19,gao2020computedram,chang.hpca16,xin2020elp2im,li.micro17,deng.dac2018,hajinazarsimdram,rezaei2020nom,wang2020figaro,ali2019memory,li.dac16,angizi2018pima,angizi2018cmp,angizi2019dna,levy.microelec14,kvatinsky.tcasii14,shafiee2016isaac,kvatinsky.iccd11,kvatinsky.tvlsi14,gaillardon2016plim,bhattacharjee2017revamp,hamdioui2015memristor,xie2015fast,hamdioui2017myth,yu2018memristive,syncron,fernandez2020natsa,cali2020genasm,kim.bmc18,ahn.pei.isca15,ahn.tesseract.isca15,boroumand.asplos18,boroumand2019conda,singh2019napel,asghari-moghaddam.micro16,DBLP:conf/sigmod/BabarinsaI15,chi2016prime,farmahini2015nda,gao.pact15,DBLP:conf/hpca/GaoK16,gu.isca16,guo2014wondp,hashemi.isca16,cont-runahead,hsieh.isca16,kim.isca16,kim.sc17,DBLP:conf/IEEEpact/LeeSK15,liu-spaa17,morad.taco15,nai2017graphpim,pattnaik.pact16,pugsley2014ndc,zhang.hpdc14,zhu2013accelerating,DBLP:conf/isca/AkinFH15,gao2017tetris,drumond2017mondrian,dai2018graphh,zhang2018graphp,huang2020heterogeneous,zhuo2019graphq,santos2017operand,ghoseibm2019,wen2017rebooting,besta2021sisa,ferreira2021pluto,olgun2021quactrng,lloyd2015memory,elliott1999computational,zheng2016tcam,landgraf2021combining,rodrigues2016scattergather,lloyd2018dse,lloyd2017keyvalue,gokhale2015rearr,nair2015active,jacob2016compiling,sura2015data,nair2015evolution,balasubramonian2014near,xi2020memory,impica,boroumand2016pim,giannoula2022sparsep,giannoula2022sigmetrics,denzler2021casper,boroumand2021polynesia,boroumand2021icde,singh2021fpga,singh2021accelerating,herruzo2021enabling,yavits2021giraf,asgarifafnir,boroumand2021google_arxiv,boroumand2021google,amiraliphd,singh2020nero,seshadri.bookchapter17,diab2022high,diab2022hicomb,fujiki2018memory,zha2020hyper,mutlu.imw13,mutlu.superfri15,ahmed2019compiler,jain2018computing,ghiasi2022genstore,deoliveira2021IEEE,deoliveira2021,cho2020mcdram,shin2018mcdram,gu2020ipim,lavenier2020,Zois2018, upmem,upmem2018,gomezluna2021benchmarking, gomezluna2022ieeeaccess, gomezluna2021cut,upmem,upmem2018,gomezluna2021benchmarking, gomezluna2022ieeeaccess, gomezluna2021cut} is a promising approach to alleviate data movement by placing processing units close to memory. As a result, NDP solutions have the potential to greatly improve system performance and energy efficiency when they are carefully designed with low-cost and low-overhead near data processing cores for memory-bound applications~\cite{Boroumand2018google}.

Our \textbf{goal} in this work is to enable high-performance and energy-efficient time series analysis for a wide range of applications, by minimizing the overheads of data movement. This can enable efficient time series analysis on large-scale systems as well as embedded and mobile devices, where power consumption is a critical constraint (e.g., heart beat analysis on a mobile medical device to predict a heart attack~\cite{LWT+19} or early earthquacke detection~\cite{Christophersen2018bayesian}). \textbf{To this end}, we propose \emph{NATSA}, the \emph{first} \underline{N}ear-Data Processing \underline{A}ccelerator for \underline{T}ime \underline{S}eries \underline{A}nalysis. The key idea of NATSA (Fig.~\ref{fig:acc_datapath}) is to exploit modern \mbox{3D-stacked} High Bandwidth Memory (HBM) along with specialized custom processing units in the logic layer of HBM, to enable energy-efficient and fast \emph{matrix profile} computation near memory, where time series data resides. NATSA supports a wide range of time series applications thanks to \emph{\di{matrix profile}}'s generality and flexibility. 

\begin{figure}[h!]
  \centering
  \includegraphics[width=0.75\linewidth]{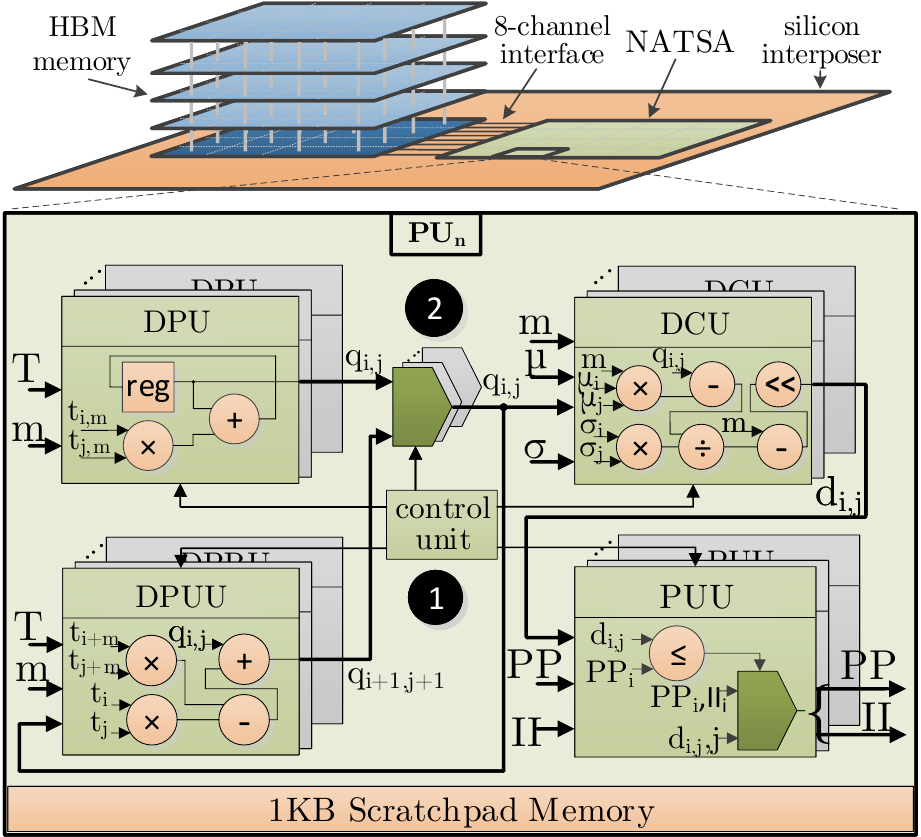}
  \caption{NATSA design and integration next to HBM memory. NATSA is connected directly to the HBM interface.}
  \label{fig:acc_datapath}
\end{figure}

Our evaluation shows that NATSA provides up to 14.2$\times$ (9.9$\times$ on average) higher performance and up to 27.2$\times$ (19.4$\times$ on average) lower energy consumption compared to a state-of-the-art multi-core system. NATSA consumes 11.0$\times$ and 4.1$\times$ less energy over optimized implementations of \emph{\di{matrix profile}} on an Intel Xeon Phi KNL~\cite{INTELKNL} and NVIDIA GTX 1050 GPU~\cite{kirk2007nvidia}, respectively. NATSA has 9.6$\times$ and 1.8$\times$ smaller area than these two accelerators, at equivalent performance points. NATSA outperforms a general-purpose NDP platform by 6.3$\times$ while consuming 10.2$\times$ less energy.

This work makes the following \textit{contributions}:
\begin{itemize}
    \item We propose \emph{NATSA}, the first near-data processing accelerator for accelerating time series analysis using modern 3D-stacked High Bandwidth Memory (HBM)~\cite{jedec.hbm.spec, lee.taco16}.
    
    \item We propose a new workload partitioning scheme that preserves the \emph{anytime} property of the algorithm, while providing load balancing among near-data processing units.
    
    \item We perform a detailed analysis of NATSA in terms of both performance and energy consumption. We compare different versions of NATSA (DDR4 and HBM) with four different architectures (8-core CPU, 64-core CPU, GPUs and NDP-CPU) and find that NATSA provides the highest performance and lowest energy consumption.
\end{itemize}

This invited extended abstract is a summary version of our prior work~\cite{fernandez2020natsa} published at ICCD 2020. NATSA's full-paper, video and codes are available at \href{https://arxiv.org/abs/2010.02079}{https://arxiv.org/abs/2010.02079}, \href{https://www.youtube.com/watch?v=PwhtSAVa\_W4}{https://www.youtube.com/watch?v=PwhtSAVa\_W4} and \href{https://github.com/CMU-SAFARI/NATSA}{https://github.com/CMU-SAFARI/NATSA}, respectively.

\end{abstract}

\section*{Acknowledgments}
This work has been supported by TIN2016-80920-R and UMA18-FEDERJA-197 Spanish projects, and Eurolab4HPC and HiPEAC collaboration grants. 
We also acknowledge support from the SAFARI Group's industrial partners, especially ASML, Facebook, Google, Huawei, Intel, Microsoft, and VMware, as well as support from the Semiconductor Research Corporation.

%%%%%%% -- PAPER CONTENT ENDS -- %%%%%%%%

\balance
%%%%%%%%% -- BIB STYLE AND FILE -- %%%%%%%%
%\bibliographystyle{unsrt}
\bibliographystyle{IEEEtranS}
\bibliography{mybibfile}
%%%%%%%%%%%%%%%%%%%%%%%%%%%%%%%%%%%%

\end{document}